\documentclass[letter]{aa} % for the letters
\usepackage[authoryear]{natbib}

\usepackage{graphicx}
\usepackage{amsmath}
\usepackage{amssymb}
\usepackage{txfonts}

\begin{document}

\title{\textsl{INTEGRAL} observation of hard X-ray variability of the TeV binary \\
LS~5039/RX~J1826.2-1450 
    \thanks{
    Based on observations with \textsl{INTEGRAL}, an ESA project with instruments
    and science data centre funded by ESA member states (especially the PI
    countries: Denmark, France, Germany, Italy, Switzerland, Spain), Czech
    Republic and Poland, and with the participation of Russia and the USA.}}

\author{A. D. Hoffmann\inst{1}, D. Klochkov\inst{1}, A. Santangelo\inst{1}, D. Horns\inst{2}, A. Segreto\inst{3}, R. Staubert\inst{1}, G. P\"uhlhofer\inst{1}}

\institute{Kepler Zentrum f\"ur Astro- und Teilchenphysik,
  Institut f\"ur Astronomie und Astrophysik, University of T\"ubingen, 
  Sand 1, 72076 T\"ubingen, Germany
  \and 
  Universit\"at Hamburg, Institut f\"ur Experimentalphysik,  Hamburg, Germany
  \and
  INAF IFC-Pa, via U. La Malfa 153, 90146 Palermo, Italy}

   \date{}

   \abstract
  % context heading (optional)
  {LS~5039\,/\,RX~J1826.2-1450 is one of the few High Mass X-ray binary
  systems from which radio and high energy TeV emission has been
  observed. Moreover, variability of the TeV emission with orbital period was detected.}  
  % aims heading (mandatory)
   {We investigate the hard X-ray ($25-200$~keV) spectral and timing properties of the source with the monitoring IBIS/ISGRI instrument on-board the \textsl{INTEGRAL} satellite. } 
  % methods heading (mandatory)
   {We present the analysis of \textsl{INTEGRAL} observations 
     for a total of about 3~Msec
     exposure time, including both public data and data
     from the Key Programme. 
     We search for flux and spectral variability related to the orbital
     phase.} 
  % results heading (mandatory)
   {The source is observed to emit from 25 up to 200~keV and the emission
     is concentrated around inferior conjunction. Orbital variability in
     the hard X-ray band is detected and established to be in phase with
     the orbitally modulated TeV emission observed with H.E.S.S. For this
     energy range we determine an average flux for the inferior conjunction
     phase interval of $(3.54 \pm 2.30) \times 10^{-11}$ erg cm$^{-2}$
     s$^{-1}$, and a flux upper limit for the superior conjunction phase
     interval of $1.45 \times 10^{-11}$ erg cm$^{-2}$ s$^{-1}$ ~(90\%
     conf. level respectively). The spectrum for the inferior conjunction
     phase interval follows a power law with an index $\Gamma =
     2.0^{+0.2}_{-0.2} $~(90\% conf. level).}
 {}  
 
 
 \authorrunning{Hoffmann, A. I. D. et al.}
 \titlerunning{LS~5039 observed by \textsl{INTEGRAL}}
 \maketitle

\section{Introduction}

LS~5039\,/\,RX~J1826.2-1450 is a High Mass X-ray Binary (HMXRB) system, which was discovered in the \emph{ROSAT} all-sky survey and identified with optical follow-up observations by \citet{motch:97}. 
The compact companion is estimated to have a mass of $M\sim3.7M_\odot$, orbiting a bright, $V=11.2$, O6.5V((f)) star with an orbital period $P_{\rm orb}\sim3.9060 \pm 0.00017$~d, an eccentricity $e=0.35\pm0.04$ and a distance of $d=2.5\pm0.1$~kpc \citep{casares:05}. 
Based on the observation of persistent asymmetric milliarcsecond radio outflows (extending up to $\sim$1000 AU), \citet{marti:98} and \citet{paredes:00} suggested the presence of a mildly relativistic ($v\sim0.2~c$) jet. \citet{paredes:02} and \citet{ribo:08} have therefore proposed the source to be a microquasar.

LS~5039 has been observed by several X-ray satellites. A summary of all
observations can be found in \citet{bosch-ramon:05a}. Flux variations on
timescales of days as well as miniflares on shorter timescales have been
observed. RXTE observations, covering the whole orbital period, show
variability with a maximum of emission at orbital phase $\phi=0.8$,
interpreted due to periastron passage. Moreover, the power-law shaped X-ray
spectrum appears harder in the high flux state \citep{bosch-ramon:05a}. 

Because inhomogeneous jets could emit $\gamma-$rays via inverse Compton scattering, \citet{paredes:00,paredes:02} have proposed the source to be the counterpart of the unidentified high energy EGRET source 3EG J1824-1514 \citep{hartman:99} and also a component of the MeV emitter GRO J1823-12, which is a superposition of three EGRET sources \citep{collmar:03}.

The source has likewise been discovered to emit at TeV energies with the
High Energy Stereoscopic System (H.E.S.S.) \citep{aharonian:05a}. This
demonstrated that X-ray binaries are capable of accelerating particles to
TeV energies. The H.E.S.S. measurements have revealed that the flux and
energy spectrum of the source are indeed modulated with the $\sim3.9$~d
orbital period of the binary system \citep{aharonian:06a}. The VHE
$\gamma$-ray emission is largely confined to half of the orbit and peaks
around the inferior conjunction (INFC) epoch of the compact companion,
where a hardening of the spectrum is observed. According to the authors
these findings indicate that $\gamma-$ray absorption with pair production
occurs in the system. Interestingly, orbital variability at TeV
$\gamma-$ray energies has also been discovered in the similar system
LS~I+61$^\circ$303 with MAGIC  \citep{albert:06} and VERITAS
\citep{acciari:08}.
 
The nature of the compact object in the system RX~J1826.2-1450\,/\,LS~5039 as well as the orgin of the TeV emission is still under debate. A microquasar scenario \citep[see e.g.][]{paredes:00,bosch-ramon:04,torres:08} as well as a pulsar scenario  \citep[see e.g][]{dubus:06,sierpowska:07,bosch-ramon:08e} have been discussed.
In this Letter, based mainly on archival data from \textsl{INTEGRAL} which span 4.6 years between March 2003 and October 2007, we confirm the detection of the source reported by \citet{bird:07} in the energy range from 25~keV up to 200~keV. We report the discovery of the modulation of the hard X-ray flux with orbital phase.

\begin{figure}
    \includegraphics[width=0.33\textwidth,angle=90.]{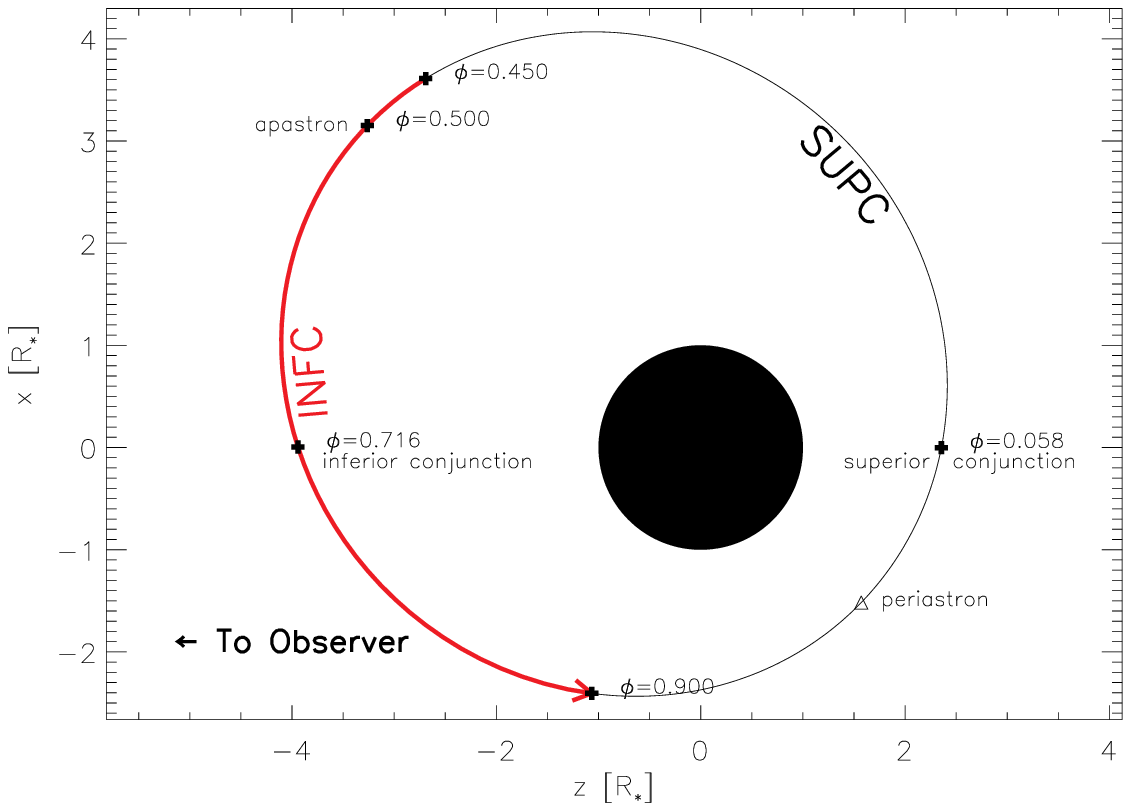}
    \caption{A sketch of the system. The orbital values are listed in \citet{casares:05} and the definition of phase intervals is from \citet{aharonian:06a}.}
    \label{fig:sketch}
\end{figure}

\section{Observations and data analysis}

\subsection{Observations}

The International Gamma-Ray Astrophysics Laboratory \citep[\textsl{INTEGRAL}, ][]{winkler:03} carries on board two main gamma-ray instruments, the \textsl{Spectrometer on INTEGRAL} \citep[SPI, 20\,keV--8\,MeV, ][]{vedrenne:03}, and the \textsl{Imager on-Board INTEGRAL Satellite} \citep[IBIS, 15\,keV--10\,MeV, ][]{ubertini:03} as well as two monitoring instruments in the X-ray and optical range, the \textsl{Joint European X-Ray Monitor} \citep[JEM-X, 3--35\,keV, ][]{lund:03}, and the \textsl{Optical Monitoring Camera} \citep[OMC, ][]{mas-hesse:03}. 

1648 science windows (ScWs) in the region that includes LS~5039 are found
in the data archive available at the \textsl{INTEGRAL} Scientific Data
Center (ISDC, http://isdc.unige.ch/), including public data as well as data
from the Key Programme. These ScWs were preselected choosing an off-axis
angle of the source smaller than 10 degrees and choosing only pointing
observations that had a nominal observation time of  more than 1000
seconds. 1627 ScWs were successfully processed through the analysis with OSA 7.0 up to the image step IMA. 

In order to reduce statistical effects a further selection of "good science
windows" was necessary. ScWs are included in the analysis when they fulfill the following criteria:

\begin{enumerate}
\item The effective exposure time after extraction is larger than $1000$~sec. 
\item The number of good time intervals (GTI) does not exceed $10$.
\item The mean background intensity level of a single ScW in the energy range $25-60$~keV falls within $3 \sigma$ of the distribution of the mean background intensity levels of all other ScWs in the same energy range.
\end{enumerate}

The resulting 1393 ScWs cover the time range from 03-11-2003 (ScW:
005000080010) to 10-25-2007 (ScW: 061400950010). Firstly, the sample
was split into two parts according to the orbital phase for imaging
analysis, and secondly into 10 smaller parts according to the orbital phase to analyse the flux dependence on orbital phase (see Sect.\ref{subsection:analysis}). Table \ref{tab:observations} summarizes the observational data of the source LS~5039 taken by the \textsl{INTEGRAL} satellite and the selection made for this analysis. 
Only in less than 10\% of the selected ScWs was LS~5039 in the fully coded field of view of JEM-X. 
Based on the sensitivity of JEM-X\footnote{The User Manual to the OSA can be found at: {\tt http://isdc.unige.ch/?Support+documents}. See p.\,15 of JEM-X Analysis User Manual for a JEM-X sensitivity plot.} and the amount of available data, a JEM-X analysis for a faint source like LS~5039 is not suitable. In this paper we therefore focus our analysis on the data of the \textsl{Integral Soft Gamma-Ray Imager} (IBIS/ISGRI).

\begin{table}[t!]
  \centering
  \caption{Summary of selected and analyzed \textsl{INTEGRAL} data for
    LS~5039. Analyzed data were taken from 69 revolutions, i.e. from revolution 0050 (03-11-2003) to revolution 0614 (10-25-2007).}
  \label{tab:observations}
  \begin{tabular}{rlc}\hline\noalign{\smallskip}\\
    Number of all extracted ScWs & 1648 & \\
    Number of ScWs after selection & 1393 & \\
    Number of ScWs in phase SUPC & 721 & \\
    Number of ScWs in phase INFC & 672 & \\  \hline
    final total analyzed exposure time & $\approx~2.99$~Msec & \\
    splitted in SUPC phase & $\approx~1.56$~Msec & \\ 
    and in INFC phase  & $\approx~1.43$~Msec & \\ \hline
  \end{tabular}
\end{table}

\subsection{Analysis}
\label{subsection:analysis}

For the analysis of ISGRI data the Off-line Science Analysis (OSA) software
(version 7.0) provided by the ISDC was used. In addition, we cross checked
the results and also generated the lightcurve in Sec.\,\ref{subsec:timing}
with the analysis package developed by INAF-IFC Palermo\footnote{\tt
  http://www.pa.iasf.cnr.it/$\thicksim$ferrigno/ INTEGRALsoftware.html}
\citep{segreto:06}.

A special handling of the OSA was used for this analysis. First, we used a user catalogue which includes 115 sources, i.e. all sources listed in the \textsl{INTEGRAL} General Reference Catalog\footnote{The \textsl{INTEGRAL} General  Reference Catalogue is available at ISDC: {\tt http://isdc.unige.ch/index.cgi?Data+catalogs}}, which are in the field of view (FoV) and which turned out to be as bright as or brighter than RX~J1826.2-1450 ($\equiv$ LS~5039).
This is required because the source of interest is quite faint and the
crowded field produces too many ghost images if the known sources are not
subtracted. Second, the spectrum was produced with the help of the OSA tool
{\sc mosaic\_spec}\footnote{\tt
  http://isdc.unige.ch/Soft/download/osa/osa\_doc/
  osa\_doc-5.0/osa\_um\_ibis-5.0/node108.html\#mosaspec} 
 from 13 mosaics, each covering one energy bin in the $13-250$~keV
  range. The binning was chosen identical to the default energy binning OSA
  uses for the spectrum calculation step (SPE). This procedure is needed due to the fact that the source is faint and the FoV is too crowded, and provides more reliable results than the standard OSA pipeline does.

We also checked our results using the standard SPE and the lightcurve
creation (LCR) steps of OSA. In this case we used a special catalogue
including all sources with a detection significance greater than $6~\sigma$
in the energy band $20-60$~keV. The catalogue includes 65
sources. According to the User Manual of the \textsl{INTEGRAL} IBIS/ISGRI
data analysis this is beyond the standard method (see the IBIS Analysis Manual\footnotemark[1] for details). However, our results show that a careful treatment of this standard method provides a comparable result in spite of the restrictions (see Fig. \ref{fig:phasogramintehess} and Sec. \ref{subsec:timing}). 

\subsubsection{Imaging} 
\begin{figure}
    \includegraphics[width=0.45\textwidth]{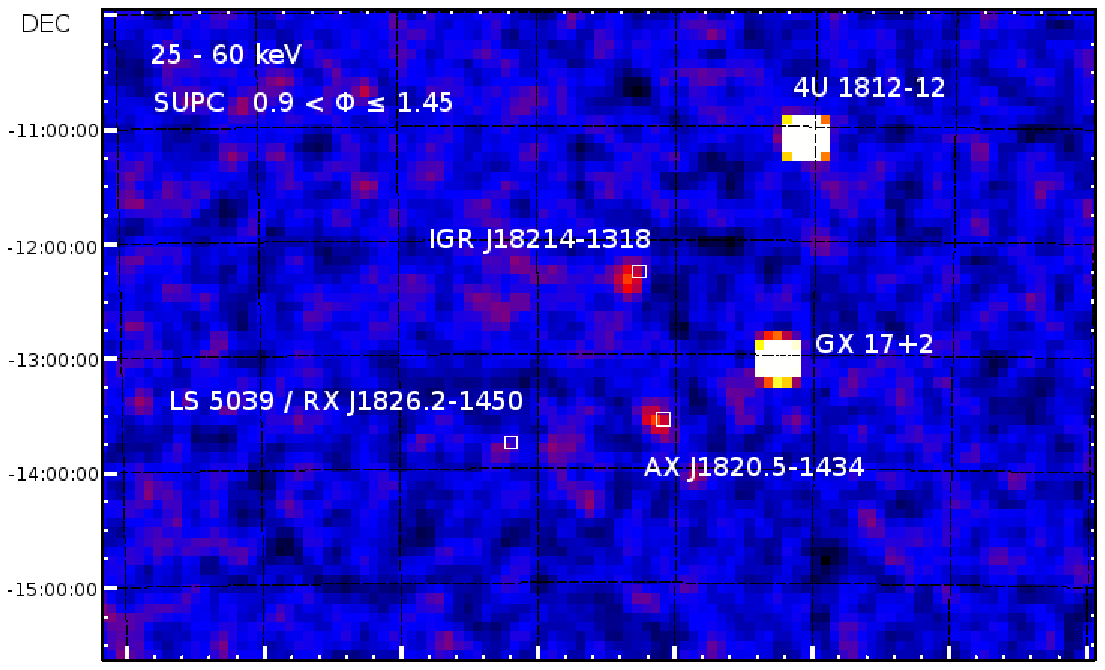}
    \includegraphics[width=0.45\textwidth]{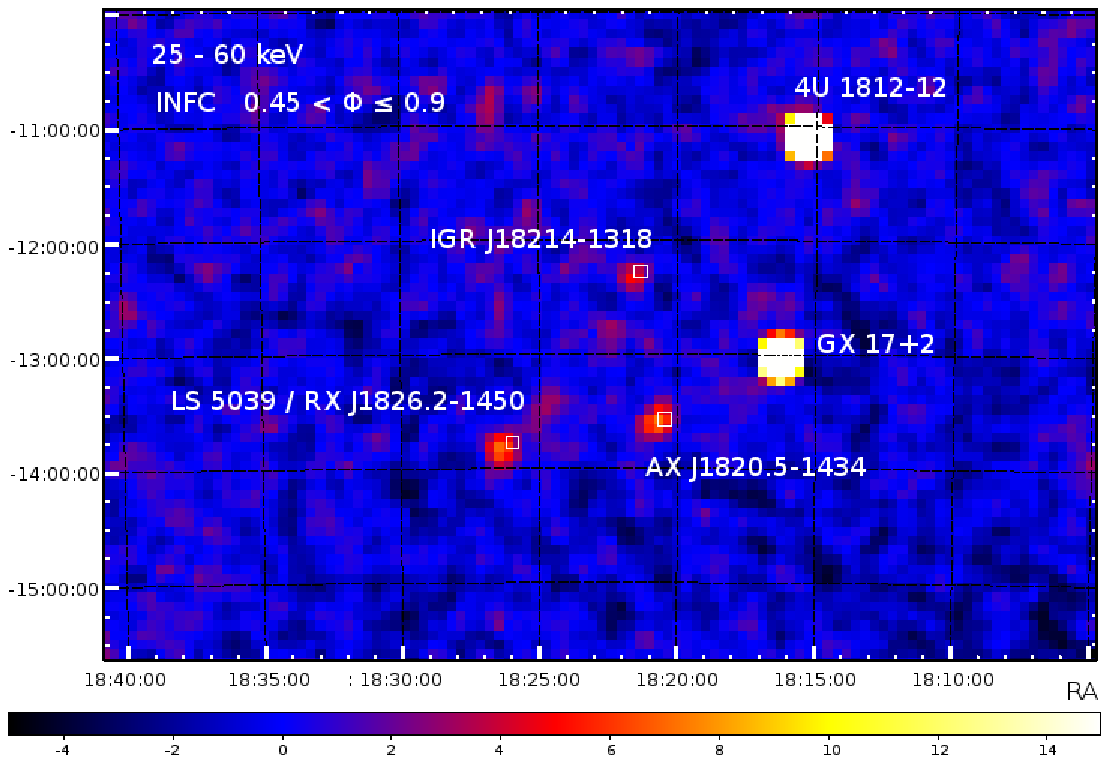}
    \caption{\textsl{INTEGRAL} significance maps in the energy range $25-60$\,keV. The open boxes are known source positions. Comparable exposure times and a higher significance for LS~5039 in the INFC phase interval indicates variability of the hard X-ray emission with orbital phase (see also Fig. \ref{fig:phasogramintehess}). The INFC phase interval includes the maximum flux state of the source. Details can be found in the text.}
    \label{fig:mosaic}
\end{figure}

The source is clearly seen at a significance level of 7.7 $\sigma$ in the
 lower energy band of $25-60$\,keV in the significance map based on the sum of
 all ScWs fulfilling the selection criteria. In addition, we divided the
 data set into two phase intervals: the superior conjunction passage at
 orbital phase $0.9 < \phi \le 1.45$ (SUPC), and the inferior
 conjunction, at orbital phase $0.45 < \phi \le 0.9$ (INFC) (see
 Figs.~\ref{fig:sketch} and \ref{fig:phasogramintehess} for details). Phase
 intervals are defined in \citet{aharonian:06a} for the H.E.S.S. data. The
 two orbital resolved significance maps in the energy range 
 $25-60$\,keV are shown in Fig.~\ref{fig:mosaic}. Positions of sources
 (open boxes) are provided by the latest \textsl{INTEGRAL} General
 Reference Catalog\footnotemark[3]. 

At the INFC phase interval, the source was detected with a significance
$\approx 7.8~\sigma$, for an exposure time of $\approx 1.43$~Msec
(Fig.~\ref{fig:mosaic}, lower panel). At the SUPC phase interval the source
was observed for a total exposure time of $1.56$~Msec, and was only
marginally detected: at the source position there is an excess at the level
of $2.8~\sigma$ (Fig.~\ref{fig:mosaic}, upper panel). For the total sample
and the total energy band $25-200$\,keV the source is detected at the
position fitted by OSA ($\rm{RA}=276.517$~deg, $\rm{DEC}=-14.812$~deg) with
$7.4 \sigma$. Within the errors, this is consistent with the source position
determinated by \citet{bird:07}. 

\subsubsection{Timing analysis} 
\label{subsec:timing}

A lightcurve produced with the INAF-IFC data analysis software
\citep{segreto:06} was used for the timing analysis. According to the
authors this software is suitable to produce reliable results even in the case of the presence of a larger number of sources in the FoV. 

\begin{figure}
  \begin{center}
    \includegraphics[width=0.33\textwidth,angle=+90.]{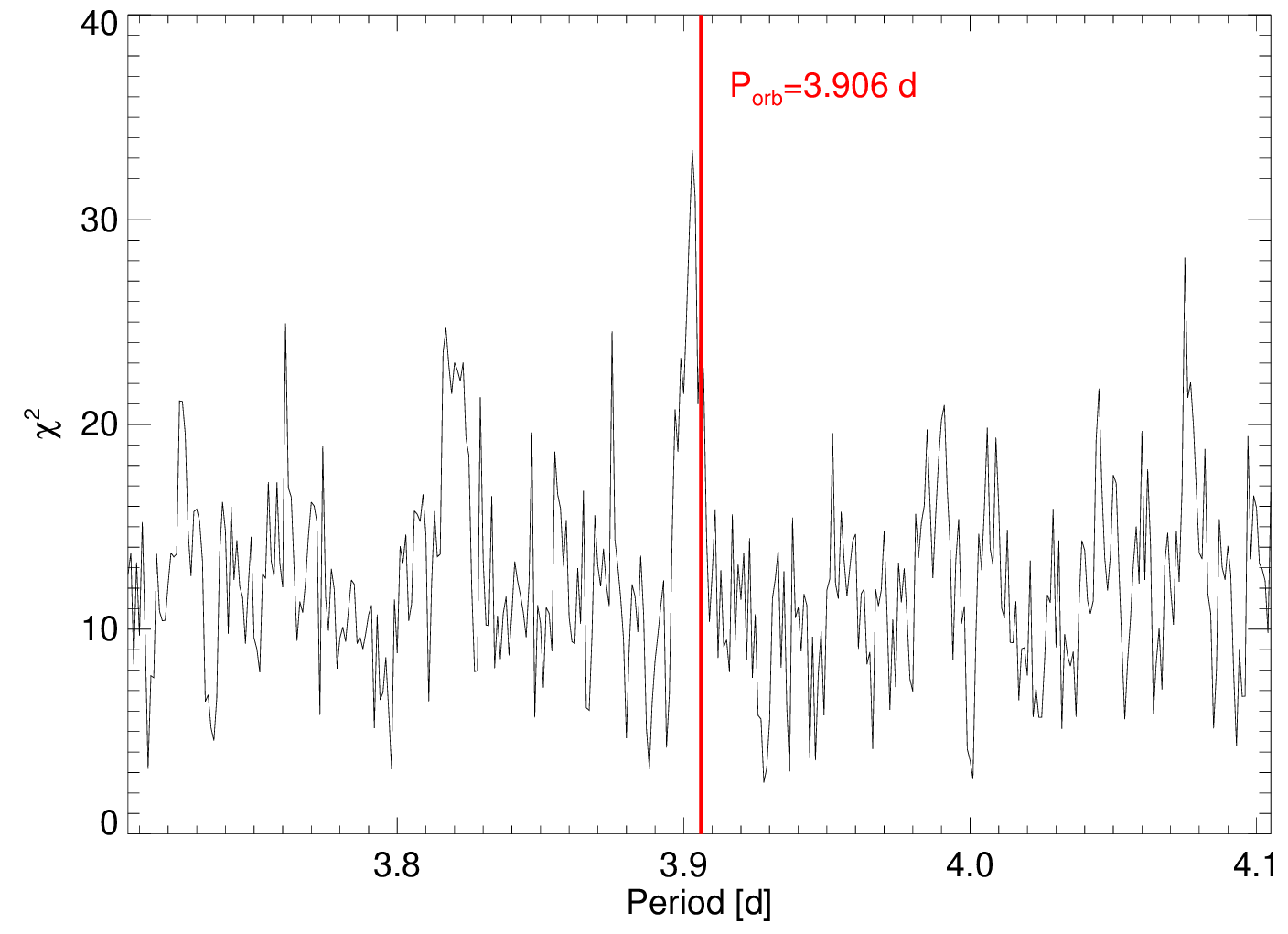}
    \caption{An epoch
      folding on the \textsl{INTEGRAL} lightcurves shows a significant
      peak at a period that is nearly identical to the orbital period known
      and determined from optical observation of the system 
      \citep{casares:05}. All other peaks are likely beat frequencies due to the large gaps in the sample of observations.}
    \label{fig:epfold}
  \end{center}
\end{figure}

Searching for an unambiguous period failed due to low statistics and large gaps in the sample of observations. But a periodogram produced by epoch folding provides a peak at a period of 3.903 days, which is within uncertainties identical to the orbital period determined by \citet{casares:05} in the optical band. Fig. \ref{fig:epfold} shows the periodogram. 

\begin{figure}
  \begin{center}
    \includegraphics[width=0.32\textwidth,angle=90.]{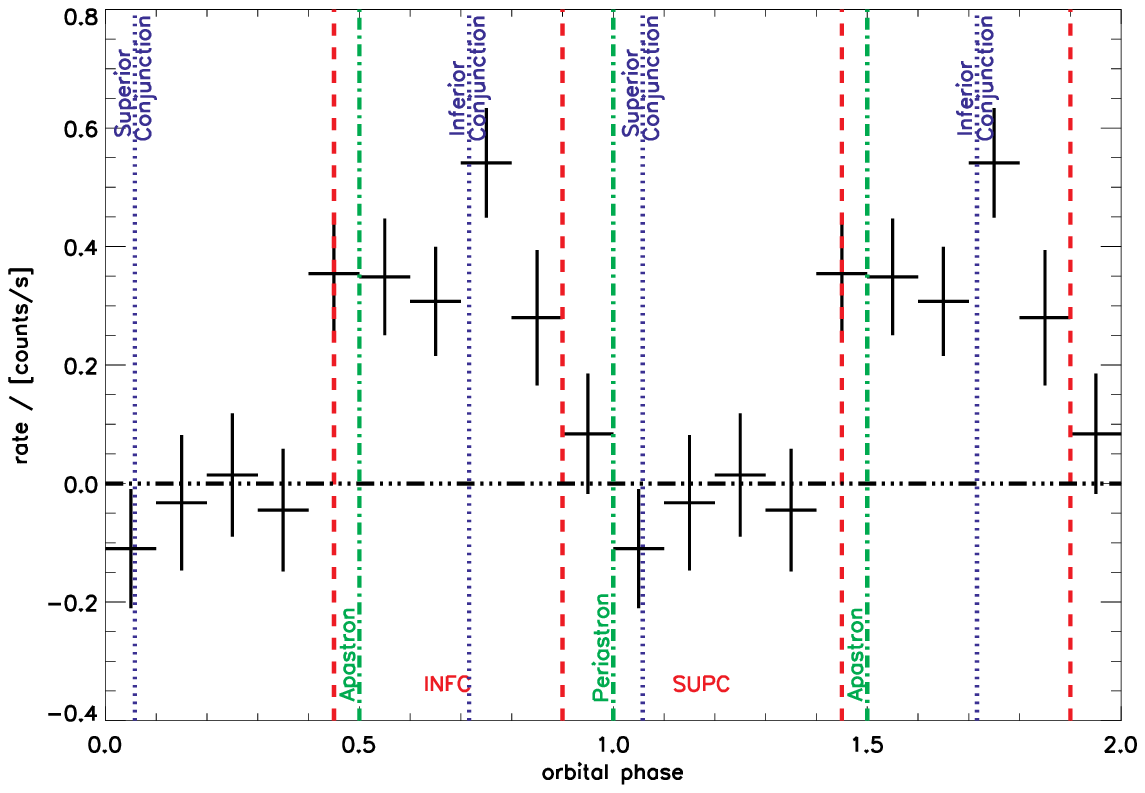}
    \includegraphics[width=0.32\textwidth,angle=90.]{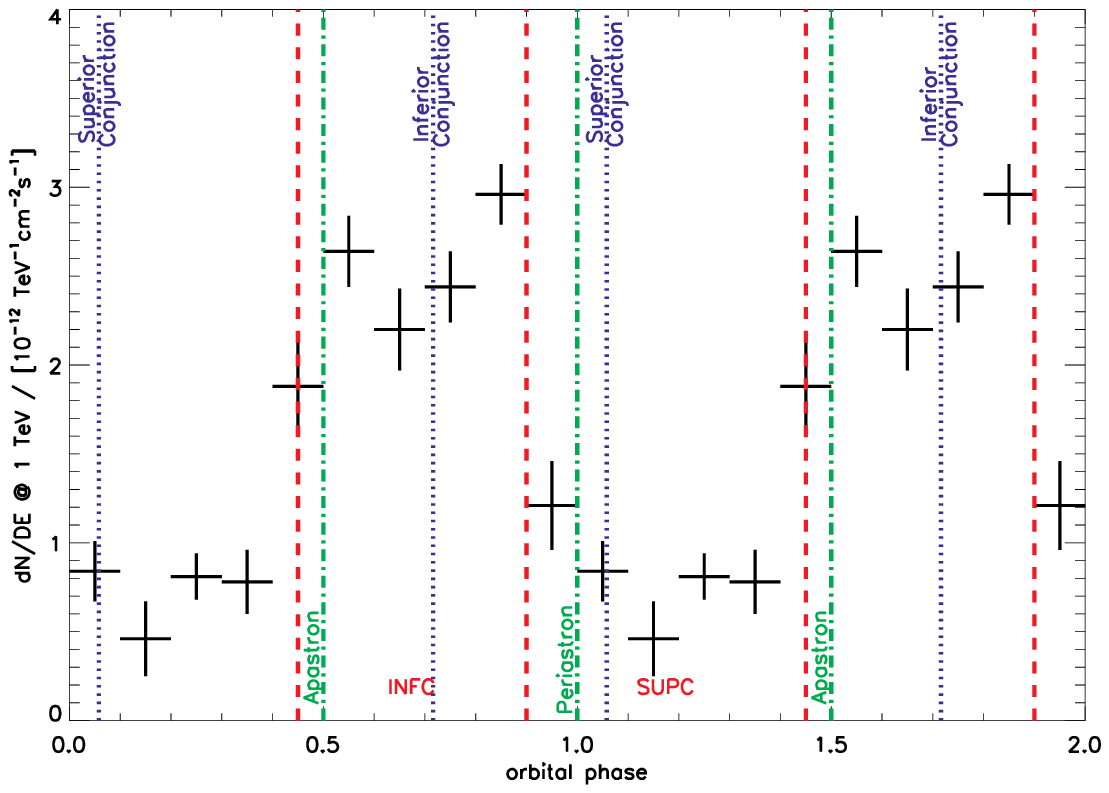}
    \caption{{\sc Top:} Orbital profile for LS~5039 in the energy range $25-200$~keV observed with \textsl{INTEGRAL}. The lightcurve is folded with the orbital period of 3.9 days. 
{Bottom:} Orbital profile for LS~5039 for the energy range above 1~TeV observed with H.E.S.S.\citep{aharonian:06a}. The hard X-ray emission is in phase with the TeV $\gamma$-rays.}
    \label{fig:phasogramintehess}
  \end{center}
\end{figure}

In a next step the lightcurve was folded using the known orbital period of $P_{\rm orb}=3.90603$ days with ${\rm HJD~}(T_{0}) = 2\,451\,943\fd09 \pm 0\fd10$ ($\equiv$~phase~$\phi=0.0$ and periastron passage) determined by \citet{casares:05}, to produce a 10 bin orbital phase profile.  
In addition, 10 mosaics corresponding to each of the 10 orbital phase intervals were produced and the count rates at the catalogue\footnotemark[3] position were determined. The values of the fluxes were derived from a 2D-Gaussian fit with fixed width (HWHM~$=6.0$~arcmin) and fixed source position. Both methods yield (within errors) comparable results:
The hard X-ray emission is clearly modulated with the orbit, and appears to be in phase with the TeV emission (Fig. \ref{fig:phasogramintehess}). The H.E.S.S. data points in the figure are from \citet{aharonian:06a}. 

\subsubsection{Spectral analysis} 
\label{sec:spectral}

A statistically significant IBIS/ISGRI spectrum of LS~5039 was obtained for the orbital phase interval corresponding to INFC (not shown here) and was modeled with a power law with $\Gamma = 2.0^{+0.2}_{-0.2}$ (90\% confidence level). The reduced $\chi ^{2}$ for the fit is 1.4 (9 dof). 
Mean countrates for the two phase intervals allow us to estimate a flux for
the INFC phase interval of $(3.54 \pm 2.30) \times~10^{-11}$
erg~cm$^{-2}$~s$^{-1}$ (90 \% conf. level). For the SUPC phase interval a flux upper limit of $1.45 \times~10^{-12}$ erg~cm$^{-2}$~s$^{-1}$ (90 \% conf. level) for the energy range $25-200$~keV can be determined.

\section{Summary and discussion}

In this paper we report on a $3$~Msec \textsl{INTEGRAL} monitoring of the HMXRB RX~J1826-1450\,/\,LS~5039. The faintness of the source and the fact that it lies in a crowded field required a very careful, non-standard handling of the data. 

Our results, obtained by applying different analysis methods, point out that the source significantly emits at hard X-rays ($25-200$~keV) and that this emission varies with the orbit in phase with the very high energy $\gamma$-rays detected with H.E.S.S. (Fig. \ref{fig:phasogramintehess}). The spectrum at the inferior conjunction is well described by a power law, while at the superior conjunction the hard X-ray emission, if any, is below the sensitivity of \textsl{INTEGRAL}.
Our result indicates that accretion might not be the mechanism for the production of the hard emission, since, in this case, we would expect a rather sharp flux maximum near periastron.

Moreover, the observed close correlation of hard X-ray and VHE $\gamma$-ray emission in
phase suggests that the emission would originate from the same source
region, possibly from particles produced in the same acceleration process.

This casts doubt on the simple explanation that the TeV variability is
mainly caused by $\gamma$-$\gamma$ absorption, as suggested in
\citet{aharonian:06a}. 

We note that, from the correlation alone, more complicated scenarios
involving $\gamma$-$\gamma$ absorption such as described by
\citet{dubus:08} cannot be ruled out. However the cascades initiated in
the $\gamma$-$\gamma$ absorption process produce secondary electrons
\citep{bednarek:97} also emitting X-ray synchrotron emission.
Therefore, soft and hard X-ray flux levels can be used to constrain the
amount of absorption involved.
The detection of TeV emission at SUPC together with the upper limit in the
hard X-ray domain derived here may pose problems for models invoking only
one emission region close to the compact object, as recently discussed in
\citet{bosch-ramon:08e}. 

Radio observations interpreted as jets have led to a microquasar interpretation for this source. In this scenario, particles are accelerated to $\approx$ TeV energies \citep{perucho:08} due to the interaction between the jet and the stellar wind of the companion. Alternatively, \citet{dubus:06} has suggested the source to be a pulsar binary system, where VHE emission arises from the pulsar wind material shocked by the interaction with the stellar companion wind and afterwards flowing away in a
comet-shape tail. Unfortunately, the observations presented here do not enable us to discriminate between the two scenarios, although the predicted X-ray variability by \citet{paredes:06} seems to be not in accordance with our observations.
Finally, we remark that a hadronic scenario \citep{romero:03,romero:08} is not
ruled out by our findings. A longer and more detailed interpretation of our
results is beyond the scope of this Letter, and will be the subject of further investigations.

\begin{acknowledgements}
We acknowledge the support of the Deutsches Zentrum f\"ur Luft- und
Raumfahrt (DLR) under grant number 50OR0302. This work is based on
observations with \textsl{INTEGRAL}, an European Space Agency (ESA)
project with instruments and science data centre funded by ESA
member states (especially the PI countries: Denmark, France,
Germany, Italy, Switzerland, Spain), Czech Republic and Poland, and 
with the participation of Russia and the USA. We also thank the anonymous
referee for his/her valueable comments. and suggestions. 
\end{acknowledgements}

\end{document}